\documentclass[aps,prl,twocolumn,superscriptaddress,nofootinbib]{revtex4-1}

\pdfoutput=1
\usepackage[normalem]{ulem}
\usepackage{graphicx}
\usepackage{amssymb,amsmath,latexsym}
\usepackage[dvipsnames]{xcolor}
\usepackage[colorlinks=true,linktocpage=true,linkcolor=blue,citecolor=orange,urlcolor=blue]{hyperref}
\usepackage{subfigure}
\usepackage{slashed}

\usepackage{soul}
\begin{document}
\title{Robust Finite-Momentum Instabilities in Dense Matter}
\author{Andr\'{e} G. da Silva} \email{andre-silva.as@acad.ufsm.br}
\affiliation{Departamento de F\'{i}sica, Universidade Federal de Santa Maria, 97105-900 Santa Maria, RS, Brazil}
\author{Ricardo L. S. Farias} \email{ricardo.farias@ufsm.br}
\affiliation{Departamento de F\'{i}sica, Universidade Federal de Santa Maria, 97105-900 Santa Maria, RS, Brazil}
\affiliation{Center for Nuclear Research, Department of Physics, Kent State University, Kent, OH 44242 USA}

\author{Theo Motta}
\email{theo.motta@unesp.br}
\affiliation{Instituto de F\'{i}sica Te\'orica, Universidade Estadual Paulista,
Rua Dr. Bento Teobaldo Ferraz, 271 - Bloco II - 01140-070 S\~ao Paulo, SP, Brazil}

\author{William R. Tavares} \email{wrtensor@gmail.com}
\affiliation{CFisUC, Department of Physics, University of Coimbra, P-3004 - 516 Coimbra, Portugal}
%
%
\begin{abstract}
The predicted extent of inhomogeneous chiral phases in effective models of quantum chromodynamics is notoriously sensitive to ultraviolet regularization. We show that this sensitivity is largely artificial. In the two-flavor Nambu--Jona-Lasinio model, conventional implementations of three-dimensional cutoff, Pauli--Villars, and proper-time regularization produce strongly different finite-momentum instability regions. Once ultraviolet regulators are restricted to genuinely divergent vacuum contributions, however, all three prescriptions yield nearly identical stability diagrams. Both the onset of the moat regime and the subsequent finite-momentum instability become quantitatively robust. The apparent scheme dependence originates from regulating ultraviolet-finite medium contributions associated with the Fermi-surface response. Our results identify spatially modulated chiral correlations as a genuine property of the dense medium rather than an artifact of the ultraviolet prescription.
\end{abstract}
\maketitle
%
%
{\it Introduction} --- The phase structure of strongly interacting matter at high baryon density remains one of the central open problems in quantum chromodynamics (QCD). While lattice simulations provide reliable information at vanishing and small baryon chemical potentials, the sign problem severely limits their applicability in the high-density regime, where functional and effective approaches continue to play a crucial role. Among the most intriguing possibilities predicted by these models is the emergence of inhomogeneous chiral phases or crystalline phases, characterized by spatially modulated chiral condensates \cite{Deryagin:1992rw,Broniowski:2011ef,Buballa:2014tba}.
The existence of such phases has been reported in a variety of frameworks, including the Nambu--Jona-Lasinio (NJL) model \cite{Rapp_2001,Ferrer:2021mpq,Basar:2009fg,Partyka:2008sv,Nakano:2004cd,Abuki:2013plajj,Fukuda:2013adajj,Hayata:2014ehajj,Moreira:2013urajj,Tatsumi:2014wkajj,Karasawa:2013zsajj,Ebert:2011rg,Frolov:2010wn,Fukushima:2008wg}, quark-meson models \cite{Rennecke:2025kub,Carignano:2014jla,Tripolt:2017zgc}, quarkyonic models \cite{Kojo_2010,Kojo:2010fe,Kojo:2011cn,McLerran:2008ua}, nuclear matter models \cite{Pitsinigkos:2023xee,Evans:2023hms,Takeda:2018ldi,Papadopoulos:2025uig,Motta:2025xop,Motta:2026pby,Lim:1989manm,Heinz:2013hzanm,Migdal:1974gdqnm,Canfora:2020uwfnm,Andersen:2018osrnm,Migdal:1978az,Dautry:1979bk}, Gross--Neveu theories \cite{Basar:2008ki,Basar:2008im,Thies:2003kk,Ciccone_2022}, generalized Ginzburg-Landau analysis and bosonic models \cite{Abuki:2011pf, Abuki:2013plajj,Iwata:2012jy,Carlomagno:2014hoa,Carlomagno:2018ogx, Schindler:2019ugo,Schindler:2021cke,Schindler:2021otf}, Dyson--Schwinger based QCD approaches \cite{Kojo_2010,Muller:2013tya,Motta:2023pks,Motta:2024agi,Motta:2024rvk}, and functional renormalization-group based QCD studies \cite{Pawlowski:2025jpg,Fu:2019hdw,Fu:2024rto,Kamikado:2012btpi}. In many cases, inhomogeneous condensates are found to occupy a significant region of the phase diagram at low temperature and moderate-to-high baryon density, often replacing the conventional first-order chiral transition. These findings suggest that spatially modulated phases may constitute a generic feature of dense strongly interacting matter.
Despite this growing evidence, the physical status of inhomogeneous chiral phases remains the subject of ongoing debate. In NJL-type models, the predicted phase structure is known to be sensitive to ultraviolet regularization procedures \cite{Broniowski:1990gb,Pannullo:2022eqh,Pannullo:2023cat,Pannullo:2023one,Pannullo:2024sov,Winstel:2024qle,Otto:2022jzl}. Since the thermodynamic potential contains vacuum contributions that require regularization, different schemes may lead to quantitatively distinct phase boundaries and, in some cases, substantially different predictions for the extent of the inhomogeneous region. This has led to the widespread perception that inhomogeneous phases may be strongly regulator dependent and could even represent artifacts of particular regularization prescriptions.
In this Letter, we revisit this long-standing issue through a systematic comparison of different regularization schemes within the NJL model. 
 Choosing a regularization procedure that reproduces physical aspects of QCD has been a non-trivial task in the context of the NJL model. Different environments involve the introduction of a new physical quantity that shifts the usual quark energy dispersion relation, which can lead to ultraviolet divergent integrations in the thermodynamic potential that mixes the cutoff with medium terms. Directly regulating such quantities can introduce several unphysical artifacts, which are typically associated with the regularization of contributions that are already finite. For such procedures we will call traditional regularization scheme (TRS). Nevertheless, despite the unavoidable aspects of such procedures, the success of an alternative method called medium separation scheme is evident. This can be verified in the case of color-superconductivity  \cite{Battistel:1998tj,Azeredo:2026wgl,XavierdeAzeredo:2026wlq,Pasqualotto:2025kpo,Azeredo:2026pbj,Casalbuoni:2003wh}, isospin unbalanced medium \cite{Lopes:2025rvn,Lopes:2021tro}, and chiral chemical potential \cite{daSilva:2025koa,Azeredo:2024sqc,Zheng:2026jqs,Liu:2025dpw}, in which the method is capable to reproduce perturbative QCD constraints for color-superconductivity and lattice QCD results for several of the remaining examples. Within this procedure, one can completely separate the usual vacuum from medium contributions.

Employing conventional approaches together with the medium separation scheme (MSS)\cite{Farias:2016let,Battistel:1998tj}, we analyze the resulting phase structure and the emergence of spatially modulated chiral condensates. We show that, once the vacuum and medium sectors are treated consistently, all three vacuum regularizations considered here lead to the same qualitative stability structure and remarkably similar quantitative results. 
Our results therefore provide strong evidence that the finite-momentum tendency toward inhomogeneous chiral symmetry breaking is not a regularization artifact, but a robust feature of the medium response in NJL-type effective descriptions of dense QCD matter. In other words, the large regulator sensitivity reported in conventional treatments can therefore be traced primarily to the regularization of ultraviolet-finite medium contributions.

In some cases, inhomogeneous phases have also been shown to be disordered by quantum and thermal fluctuations \cite{Pisarski:2018bct,Winstel:2024qle,Lee:2015bva,Hidaka:2015xza}. What remains, however, is not completely trivial. Although the phase is disordered, correlation functions still manifest  a remnant oscillatory behavior typical of inhomogeneous phases. These are oscillatory and decaying static correlation functions which correspond in momentum space to a zero-frequency correlation whose minimum occurs at nonzero spatial momentum \cite{Pisarski:2021qof,Nussinov:2024erh,Rennecke:2023xhc}. This behaviour of the correlation functions defines the so-called moat regime. As opposed to inhomogeneous phases, these have been reported to be at least qualitatively robust under regulator changes \cite{Pannullo:2024sov}. As we will show, once convergent medium contributions are kept unregularized within the MSS, this qualitative robustness becomes quantitative. Both the moat regime and the finite-momentum instability become remarkably insensitive to the choice of vacuum regulator.

{\it Model and Analysis} --- We consider the two-flavor NJL model in the chiral limit. The bosonized Lagrangian is written as
\begin{equation}
    \mathcal{L} = \bar{\psi} \left[ i \slashed{\partial} - \sigma - i\gamma_5 \vec{\tau} \cdot \vec{\pi} \right] \psi 
    - \frac{\sigma^2 + \vec\pi^2}{4 G},
\end{equation}
where $\psi(x)$ is the quark field with flavor and color degrees of freedom, $\sigma({x})$ and $\vec{\pi}({x})$ are the auxiliary bosonic fields, $\vec{\tau}$ are the Pauli matrices acting on isospin space, and $G$ is the coupling constant with units of GeV$^{-2}$ (in natural units). In mean-field approximation, the auxiliary fields are replaced by their expectation values, which are proportional to the quark bilinear condensates. 

Since the action is quadratic in $\psi$, we can now integrate out the fermion fields and obtain the Landau free energy of the system as a functional of the classical auxiliary fields. Equilibrium configurations are obtained as stationary points of the free energy, however, the completely symmetric solution, where $\sigma=\pi_a=0$ is always an equilibrium configuration. We would like to analyse the stability of the symmetric solution against spatially inhomogeneous breaking of chiral symmetry, therefore, we expand the free energy around infinitesimally small perturbations of the $\sigma$ field, i.e. $\delta\sigma(\boldsymbol{x})$.
Expanding the effective action to quadratic order in the perturbations shows that the fully symmetric solution is locally stable when all eigenvalues of the static two-point function $\Gamma^{(2)}(\boldsymbol{q}^2)$ are nonnegative. If negative eigenvalues appear only at nonzero tri-momenta, it signals an instability toward spatial modulation \cite{Buballa:2014tba}. 
It is useful to distinguish the onset of a finite-momentum minimum from the actual loss of stability. 
We define the moat regime as the region in which the second derivative of the two-point function at zero momentum, which is proportional to the wave-function renormalization, is negative~\cite{Pisarski:2021qof,Pannullo:2024sov},
\begin{equation}
    \left. \frac{\partial^2 \Gamma^{(2)}}{\partial q^2} \right|_{q = 0} < 0,
\end{equation}
whereas a genuine finite-momentum instability is signaled by
\begin{equation}
    q_{\rm min} \neq 0,
    \qquad
    \Gamma^{(2)}(q_{\rm min}) < 0.
\end{equation}
Thus, the moat regime should be understood as a precursor of spatial modulation: the homogeneous phase might remain stable, but its softest static mode already carries a finite momentum. 
With these considerations, the static two-point function is given by
\begin{equation}
    \Gamma^{(2)}(\boldsymbol{q}^2) = \frac{1}{2G} - 4 N_f N_c \ell_1 + 4 N_f N_c \ell_2(\boldsymbol{q}^2),
\end{equation}
where $E_{\boldsymbol{p}} = |\boldsymbol{p}|$,  $N_f = 2$ and $N_c = 3$. The gap equation integral $\ell_1$ and the $\boldsymbol{q}^2$-dependent integral $\ell_2$ are given by
\begin{equation}
\begin{aligned}
    \ell_1 = \int \frac{d^3p}{(2\pi)^3} \frac{1}{E_{\boldsymbol{p}}}\left[ 1 - n(E_{\boldsymbol{p}}) - \bar{n}(E_{\boldsymbol{p}}) \right], \\ 
    \ell_2(\boldsymbol{q}^2) = \int \frac{d^3p}{(2\pi)^3} \frac{\boldsymbol{p} \cdot \boldsymbol{q} + \boldsymbol{q}^2 }{2 \boldsymbol{p} \cdot \boldsymbol{q} + \boldsymbol{q}^2} \bigg[ \frac{1 - n(E_{\boldsymbol{p}}) - \bar{n}(E_{\boldsymbol{p}})}{2E_{\boldsymbol{p}}} \\
    - \frac{1 - n(E_{\boldsymbol{p} + \boldsymbol{q}}) - \bar{n}(E_{\boldsymbol{p} + \boldsymbol{q}})}{2E_{\boldsymbol{p} + \boldsymbol{q}}} \bigg],
\end{aligned}
\end{equation}
where $n(E_{\boldsymbol{p}})$ and $\bar{n}(E_{\boldsymbol{p}})$ are Fermi-Dirac distributions. While $\ell_1$ contains a divergent vacuum contribution, it is independent of the external momentum $\boldsymbol{q}$. Its divergent vacuum part can therefore be isolated and regularized independently of the finite medium contribution. Conversely, $\ell_2(\boldsymbol{q}^2)$ contains a divergent vacuum contribution that depends on the external momentum and is given by
\begin{equation}
    \ell_{2}^{\text{vac}}(\boldsymbol{q}^2) = \int \frac{d^3p}{(2\pi)^3} \frac{\boldsymbol{p} \cdot \boldsymbol{q} + \boldsymbol{q}^2 }{2 \boldsymbol{p} \cdot \boldsymbol{q} + \boldsymbol{q}^2} \left( \frac{1}{2E_{\boldsymbol{p}}} - \frac{1}{2E_{\boldsymbol{p} + \boldsymbol{q}}} \right).
\end{equation}

To obtain physically consistent results, however, one must avoid regularizing contributions that are already ultraviolet finite, since doing so can introduce an artificial dependence on the regularization procedure and scale.


{\it Medium Separation Scheme} ---  
 The NJL model is nonrenormalizable and therefore requires an ultraviolet prescription for divergent vacuum integrals. In conventional finite-density calculations, the regulator is often applied directly to the complete momentum integral, thereby also acting on ultraviolet-finite medium contributions. The medium separation scheme reorganizes the integrals so that all divergent terms are expressed in terms of vacuum quantities, while the remaining medium-dependent contributions are kept unregularized. In the present problem, this separation must also preserve the finite dependence on the external momentum that controls the stability against spatially modulated fluctuations.

To extract the finite contributions, we start from the momentum-shifted form of $\ell_{2}^{\text{vac}}$
\begin{equation}
    \ell_{2}^{\text{vac}}(\boldsymbol{q}^2) = \int \frac{d^3 p}{(2\pi)^3} \frac{\boldsymbol{q}^2}{4 \boldsymbol{p} \cdot \boldsymbol{q}} \left( \frac{1}{2 E_{\boldsymbol{p} - \boldsymbol{q}/2}} - \frac{1}{2 E_{\boldsymbol{p} + \boldsymbol{q}/2}} \right).
\end{equation}
Using the proper-time representation of this integral, we identify a divergent contribution that is independent of $\boldsymbol{p} \cdot \boldsymbol{q}$. By adding and subtracting this contribution, we isolate it from the finite remainder,
\begin{equation}
\begin{aligned}
    \ell_{2}^{\text{vac}}&(\boldsymbol{q}^2) = \frac{\boldsymbol{q}^2}{32\pi^{2}} \int \frac{d\tau}{\tau} e^{-\tau\boldsymbol{q}^2/4} \\
    &+ \frac{|\boldsymbol{q}|}{16\pi^{5/2}} \int_{0}^{\infty} d\tau \; \tau^{1/2} e^{-\tau\boldsymbol{q}^2/4} \\
    &\times \int_{-1}^{1} dx \int_{0}^{\infty} dp \frac{p}{x} e^{-\tau p^2} \left[ \sinh(\tau p q x) - \tau p q x \right].
\end{aligned} 
\end{equation}
Performing the integration of the second term we arrive at the separated integral
\begin{equation}
    \ell_{2}^{\text{vac}}(\boldsymbol{q}^2) = \frac{\boldsymbol{q}^2}{32\pi^2}\left[ I^{(\log)}(\boldsymbol{q}^2/4) + 2 - \ln 4 \right],
\end{equation}
where
\begin{equation}
    I^{(\log)}(\boldsymbol{q}^2/4) = \int_\Lambda \frac{d\tau}{\tau} e^{-\tau \boldsymbol{q}^2/4}
\end{equation}
is a logarithmically divergent integral to be regulated via three-dimensional cutoff, Pauli-Villars regularization, and proper-time regularization.
It is important to note that not all of the $\boldsymbol{q}$ dependence can be extracted from the regulated integral. Performing an additional extraction leads to both an infrared divergence in the chiral restored phase and a $-\boldsymbol{q}^2\ln(\boldsymbol{q}^2/\Lambda^2)$ dependence, making the two-point function pathologically negative at $|\boldsymbol{q}| \gg \Lambda$.

\begin{figure*}
    \centering
    \includegraphics[width=0.9\linewidth]{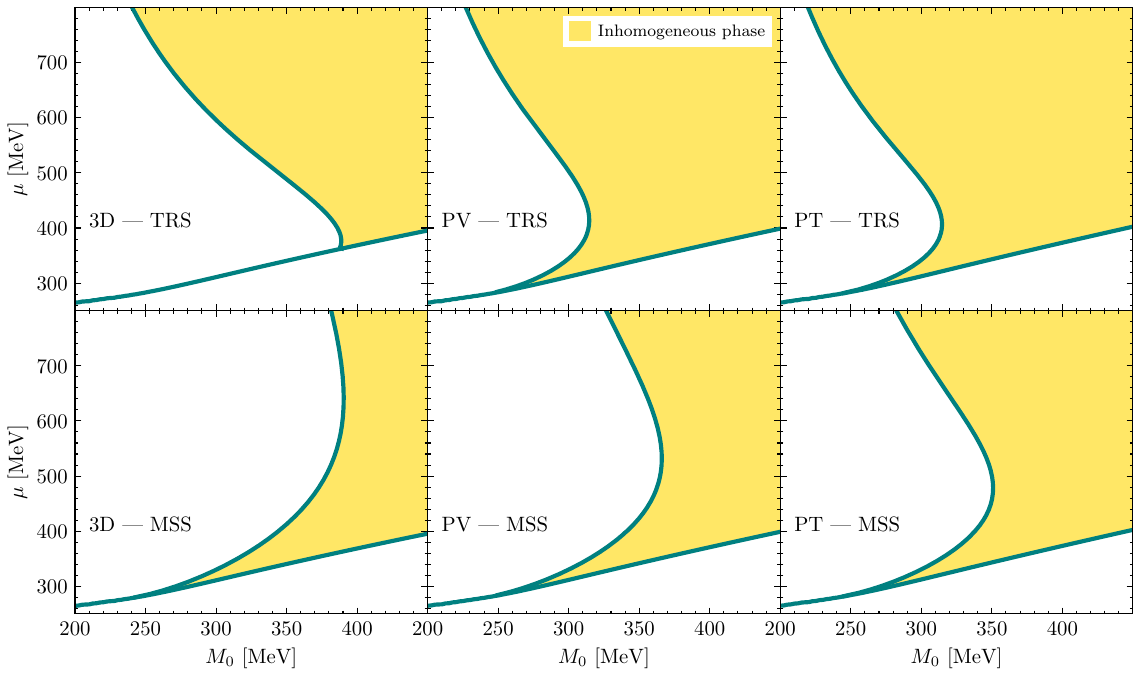}
    \caption{
Regions of inhomogeneous instability in the $(M_0,\mu)$ plane obtained from the stability analysis of the homogeneous solution. The upper and lower panels show results obtained with the traditional regularization scheme (TRS) and the medium separation scheme (MSS), respectively, using a three-dimensional momentum cutoff (3D), Pauli--Villars regularization (PV), and proper-time regularization (PT). The shaded regions correspond to $\min_{\boldsymbol{q}}\Gamma^{(2)}(\boldsymbol{q}^2)<0$.}
    \label{fig1}
\end{figure*}

Although the NJL model is nonrenormalizable, this does not imply that all finite-density observables inherit a comparable regulator ambiguity. Once the divergent vacuum sector is separated from ultraviolet-finite medium contributions, the finite-momentum response can remain quantitatively stable across distinct ultraviolet prescriptions.

A useful physical interpretation of the separation becomes particularly transparent at finite density. The momentum dependence of $\Gamma^{(2)}(\boldsymbol{q}^2)$ that drives the moat regime and eventually the finite-momentum instability receives essential contributions from medium excitations. These contributions are ultraviolet finite and encode the response of states in the vicinity of the Fermi surface. Applying a vacuum regulator to the complete momentum integral therefore distorts precisely the finite-density dynamics responsible for the instability. Different regulator kernels modify this response in different ways, thereby producing an artificial scheme dependence. In the MSS, by contrast, only the genuinely divergent vacuum structures retain regulator dependence, whereas the finite medium response remains intact.

{\it Phase diagram} --- Figure~\ref{fig1} exposes the origin of the apparent regularization ambiguity. Within the traditional regularization scheme, the instability domain in the ($M_0,\mu$) plane , where $M_0$ is the quark mass in the vacuum (i.e., evaluated at $T=\mu=0$), changes substantially among the three-dimensional cutoff, Pauli--Villars, and proper-time prescriptions, displaying sizable shifts and even qualitatively different turning structures. This sensitivity is largely removed by the MSS: once the regulator is restricted to genuinely divergent vacuum contributions, the three prescriptions yield nearly overlapping instability domains over the physically relevant range of $M_0$. The comparison therefore indicates that the large scheme dependence found in conventional calculations is not an intrinsic uncertainty of the finite-momentum instability, but is predominantly generated by regulating ultraviolet-finite medium contributions that govern the response near the Fermi surface\footnote{For each value of $M_0$, the coupling and regulator scale are refitted to the corresponding vacuum observables.}.

\begin{figure*}
    \centering
    \includegraphics[width=0.9\linewidth]{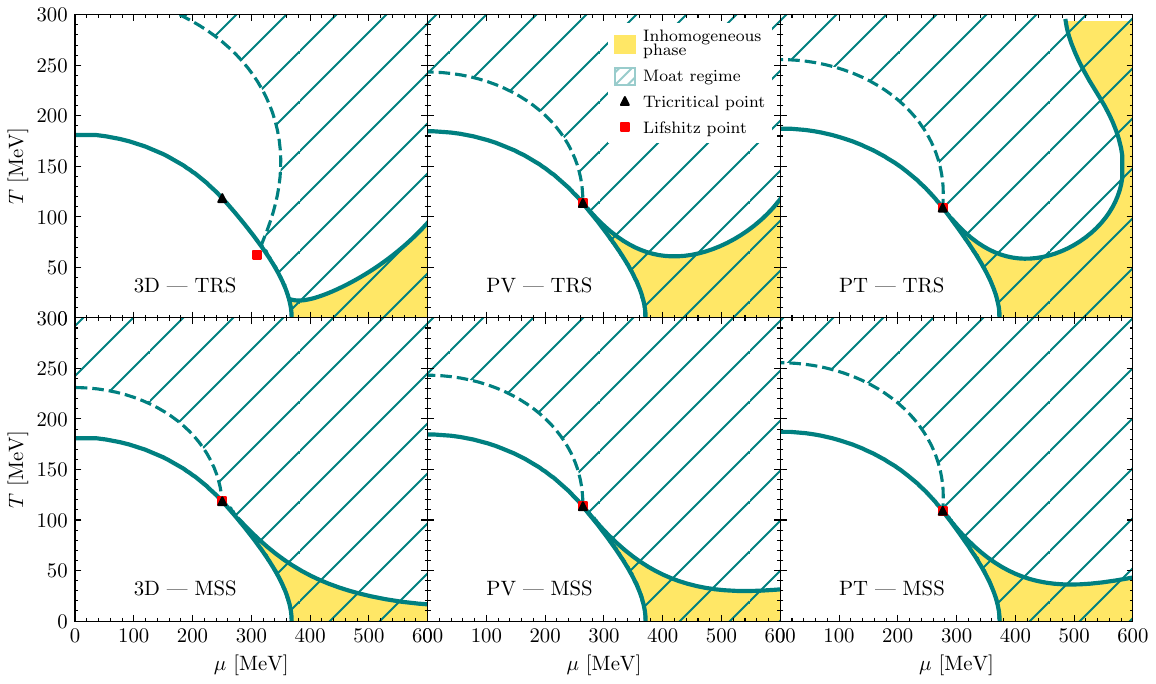}
    \caption{
Phase diagrams in the $(T,\mu)$ plane for $M_0=400~\mathrm{MeV}$, obtained from the stability analysis of the homogeneous solution. The upper and lower panels show results obtained with the traditional regularization scheme (TRS) and the medium separation scheme (MSS), respectively, using three-dimensional momentum cutoff (3D), Pauli--Villars (PV), and proper-time (PT) regularizations. The shaded regions satisfy $\min_{\boldsymbol{q}}\Gamma^{(2)}(\boldsymbol{q}^2)<0$ and indicate an instability toward an inhomogeneous condensate. The dashed curves mark the onset of the moat regime, where the minimum of the static two-point function moves to nonzero spatial momentum. Black triangles and red squares denote the tricritical and Lifshitz points, respectively.}
    \label{fig2}
\end{figure*}

The same conclusion becomes even more striking in the $T-\mu$ phase diagrams of Fig.~\ref{fig2}. Traditional prescriptions yield mutually incompatible high-density behavior, including strongly regulator-dependent reentrant instability regions and substantial variations in their thermal extent. In contrast, the MSS produces a common phase structure for all three vacuum regularizations: the inhomogeneous-instability boundaries nearly coincide, and the onset of the moat regime, marked by the dashed curves, becomes quantitatively stable. Remarkably, the MSS moat boundary bends toward lower chemical potential as the temperature increases, reproducing the qualitative curvature found independently in the quark-meson model of Rennecke and Yin~\cite{Rennecke:2025kub}. 
This contrasts with the characteristic curvature found in conventionally regularized NJL calculations \cite{Pannullo:2024sov}, suggesting that part of this difference may originate from the regularization of ultraviolet-finite medium contributions.
The qualitative agreement of the MSS moat curvature with quark-meson-model results is particularly suggestive, since the two approaches differ substantially in their treatment of ultraviolet physics. This points to the shape of the moat boundary being controlled primarily by the finite-density response rather than by model-specific ultraviolet details. Thus, the robustness extends beyond the region where the homogeneous solution is already unstable; it also encompasses the finite-momentum precursor regime in which the static two-point function develops its global minimum at nonzero momentum. This separation between the onset of a finite-momentum minimum and the actual loss of stability is physically important. It shows that the tendency toward spatial modulation is already encoded in the medium response before the homogeneous phase becomes unstable. The regulator robustness of the moat regime therefore indicates that the finite-momentum structure is not merely a property of the eventual instability boundary, but a genuine precursor phenomenon. These results show that both the moat regime and the subsequent inhomogeneous instability are controlled by medium dynamics and become insensitive to the ultraviolet prescription once vacuum and medium contributions are consistently separated. The moat boundary identifies the point where the global minimum of the static inverse propagator shifts from $q=0$ to $q\neq0$, whereas the Lifshitz point marks the meeting of homogeneous and modulated instability lines.

The near collapse of the MSS curves is therefore nontrivial: the three prescriptions still regulate the divergent vacuum sector differently, but become equivalent at the level of the finite-density mechanism responsible for the finite-momentum response.

{\it Final remarks} --- 
We have shown that the pronounced regulator dependence of finite-momentum instabilities in the two-flavor NJL model is largely generated by applying ultraviolet regularization to contributions that are finite and medium dependent. Traditional prescriptions lead to qualitatively different instability domains and moat boundaries, whereas the medium separation scheme yields a common phase structure for three-dimensional cutoff, Pauli--Villars, and proper-time vacuum regularizations. The robustness of both the finite-momentum minimum and the subsequent loss of stability demonstrates that these structures are controlled primarily by the medium response rather than by the ultraviolet prescription. Our analysis establishes the local onset of spatial modulation, while determining the fully developed inhomogeneous ground state and its fluctuation corrections remains an important next step. 

More broadly, these results indicate that finite-momentum structures in dense effective theories can remain quantitatively predictive even in nonrenormalizable frameworks, provided ultraviolet and medium physics are consistently disentangled.

{\it Acknowledgments} --- This work was partially supported by Conselho Nacional de Desenvolvimento Cient\'ifico e Tecno\-l\'o\-gico  (CNPq), Grants No. 312032/2023-4, No. 402963/2024-5 and 445182/2024-5 (R.L.S.F.),  200037/2026-9 (W.R.T.) and 315225/2025-4 (TFM); Funda\c{c}\~ao de Amparo \`a Pesquisa do Estado do Rio 
Grande do Sul (FAPERGS), Grants No. 24/2551-0001285-0 (R.L.S.F.), No. 23/2551-0000791-6 and No. 23/2551-0001591-9 (D.C.D.). CAPES Finance Code 001 (A.G.S). The work is also part of the project
Instituto Nacional de Ci\^encia e Tecnologia - F\'isica Nuclear e
Aplica\c{c}\~oes (INCT - FNA), Grants No. 464898/2014-5 and No. 408419/2024-5, and supported
by the Ser\-ra\-pi\-lhei\-ra Institute (Grant No. Serra -
2211-42230). R. L. S. F. acknowledges
the kind hospitality of the Center for Nuclear Research at Kent State University, where part of this work was
done. W.R.T. is grateful to the Centro de Física at the Universidade de Coimbra for its hospitality.
%
\bibliography{refs}
%

\end{document}